\documentclass[preprint,aps]{revtex4}

\usepackage{graphicx}

\begin{document}


\title{Semimetalic graphene in a modulated electric potential}

\author{J. H. Ho , Y. H. Chiu, S. J. Tsai}
\author{M. F. Lin}%
 \email{mflin@mail.ncku.edu.tw}
\affiliation{%
Department of Physics, National Cheng Kung University, Tainan,
Taiwan 701
}%

\date{\today}

\begin{abstract}
The $\pi$-electronic structure of graphene in the presence of a
modulated electric potential is investigated by the tight-binding
model. The low-energy electronic properties are strongly affected by
the period and field strength. Such a field could modify the energy
dispersions, destroy state degeneracy, and induce band-edge states.
It should be noted that a modulated electric potential could make
semiconducting graphene semimetallic, and that the onset period of
such a transition relies on the field strength. There exist infinite
Fermi-momentum states in sharply contrast with two crossing points
(Dirac points) for graphene without external fields. The finite
density of states (DOS) at the Fermi level means that there are free
carriers, and, at the same time, the low DOS spectrum exhibits many
prominent peaks, mainly owing to the band-edge states.
\end{abstract}

\pacs{73.20.At, 73.22.-f, 81.05.Uw}%
\maketitle

\section{Introduction}
Carbon atoms could form diamond, graphites, carbon nanotubes,
C60-related fullerenes, carbon onions, and carbon tori. These
systems own special symmetric configurations, and their
dimensionalities range from 3D to 0D. These systems are of great
interest to community because of the spectacular physical properties
to which the unique geometric structures give rise. Recently, a new
carbon-based material is discovered by the success in experimental
realization of fabricating a single graphene sheet
\cite{discov-1,discov-2}, and immediately this unique 2D system
becomes the focus of both theoretical and the experimental
researches. One of the unique physical properties of the kind of
system, for example, is the observation of an anomalous quantum Hall
effect \cite{Mless}, where the plateau of zero Hall conductance is
unexpectedly absent. The novel phenomenon of Hall conductivity
quantization is attributed to that the dynamics of quasiparticles in
graphene is effectively relativistic, contrasting sharply with the
conventional integer quantum Hall effect in the regular 2D electron
system that are governed by the Schrodinger fermions.

Graphene has a honeycomb crystal structure with two distinct
triangular sublattices A and B. This real-space structure
corresponds to a triangular lattice in the reciprocal space, which
also defines a hexagonal Brioullin zone with two inequivalent
corners, $K$ and $K'$. The unique geometrical configuration of
graphene leads to several non-trivial characteristics of its energy
spectrum. There exist two identical low-energy bands in the vicinity
of the two inequivalent corners, of which each exhibits a linear
energy dispersion with conduction and valence bands crossing right
at the corner. The highly diminished Fermi surface gives rise to a
zero density of states at the Fermi level, so graphene a zero-gap
semiconductor.

Studying the behavior of electrons under various kinds of external
fields is not only of vital importance for understanding their
physical nature of materials but also helpful for designing novel
devices or developing applications. Many researches has been
conducted on the physical properties of graphene, such as electronic
properties \cite{band-1,band-2,band-3,band-4,band-5}, transport
properties
\cite{discov-1,transp-1,transp-2,transp-3,transp-4,transp-5},
optical properties \cite{opt-1,opt-2,opt-3,opt-4,opt-5,opt-6}, and
electronic excitations \cite{excit-1,excit-2,excit-3}. Here we focus
on the effects of modulated fields on the electronic properties, an
intriguing field of less exploration
\cite{mod-1,mod-2,mod-3,mod-4,modB:jon,modL:jon}. The systems for
graphene in the modulated fields are not yet realized
experimentally; however, the 2D electron gas formed by GaAs/AlGaAs
heterojunctions in modulated fields are well-established systems and
have been subject of active studies in the past two
decades\cite{modE:weiss,modE-1:exp,modB-1:exp,modB-2:exp}. For the
2D electron gas, there are probably two types of periodic
modulations in actual systems. One is an electrostatic modulation
and the other is a magnetic-field modulation. The former can be
realized by a periodic array of gate electrodes \cite{modE-1:exp}
and the latter by depositing magnetic materials on the surface
\cite{modB-1:exp,modB-2:exp}. In the case of magnetic modulation,
the electronic properties of graphene have been recently studied
\cite{modB:jon,modL:jon}. In this paper, we would like to further
investigate the electronic properties of graphene in a modulated
electric potential. The tight-binding model is employed to calculate
the energy spectrum and the density of states. Both are analysed as
functions of the field strength, the period, and the direction by
solving the Hamiltonian matrix numerically. In this work, it is
shown that semiconducting graphene can be made semimetallic by
applying a modulated electric potential. Any finite value of field
strength will cause such a transition with a prerequisite that the
period of modulated potential is longer enough. However, a
qualitative study, recently made by other authors, on the Dirac
particles tunneling through a one-dimensional potential barrier
predicts that the 2D light-cone structure is still preserved under
such a field \cite{mod-4}. The underlying reason causing the
different findings is elaborated in the discussion part.

This paper is organized as follows. The tight-binding Hamiltonian
matrix in a periodic electric potential is derived in Sec. II. The
main characteristics of the $\pi$-electronic structures are
discussed in Sec. III. Finally, Sec. IV contains concluding remarks.

\section{Model and Methods: Tight-binding Model}
The $\pi$-electronic structure of graphene is calculated by the
tight-binding model within the nearest-neighbor interactions. Given
that graphene is a two-dimensional triangular lattice with two-point
basis, we construct the $\pi$-electronic eigenfunction of the
system:
\begin{eqnarray}
|\Psi_{\mathbf{k}}\rangle=C^A_{\mathbf{k}}|\Phi_{\mathbf{k}}^A\rangle+C^B_{\mathbf{k}}|\Phi^B_{\mathbf{k}}\rangle,
\end{eqnarray}
where the tight-binding Bloch function
${|\Phi^A_{\mathbf{k}}\rangle}$ ($|\Phi^B_{\mathbf{k}}\rangle$) is
the superposition of the $2p_z$ orbitals from periodic $A$ ($B$)
atoms. The coefficients are obtained by diagonalizing the
Hamiltonian matrix in the space spanned by the tight-binding Bloch
functions. Moreover, with the two prerequisites that site energies
for A and B atoms are the same, and that intra-sublattice hoppings
are ignored, we can set
$\langle\,\Phi^A_{\mathbf{k}}|H_0|\Phi^A_{\mathbf{k}}\rangle
=\langle\,\Phi^B_{\mathbf{k}}|H_0|\Phi^B_{\mathbf{k}}\rangle\,=0$.
The nonvanishing matrix elements are given by
\begin{eqnarray}
\langle\,\Phi^B_{\mathbf{k}}|H_0|\Phi^A_{\mathbf{k}}\rangle
=\frac{1}{N}\sum_{\mathbf{T}_i}\gamma_0\exp\{-i\mathbf{k}\cdot
\mathbf{T}_i\}\equiv\sum_i t_{i\mathbf{k}},\; i=1, 2, 3,
\end{eqnarray}
where
$\gamma_0\equiv\int\varphi_{p_z}^\ast(\mathbf{r}-\mathbf{R}_A)H_0
\varphi_{p_z}(\mathbf{r}-\mathbf{R}_A-\mathbf{T}_i)d^3r\simeq 2.6$
eV is the nearest-neighbor hopping integral with $R_A$ being the
lattice vector and $\varphi_{p_z}$ the atomic $2p_z$ orbital.
$\mathbf{T}_i$ are the vectors connecting a carbon atom to its three
nearest neighbors, and they are given by
$\mathbf{T}_1=(b/2,\sqrt{3}b/2)$, $\mathbf{T}_2=(b/2,-\sqrt{3}b/2)$,
and $\mathbf{T}_3=(-b,0)$, where $b=1.42$ $\mathring{A}$ is the C-C
bond length. $t_{i\mathbf{k}}$, denoting individual hopping process,
is then explicitly given by
$t_{1\mathbf{k}}=\gamma_{0}\exp[\,(ik_{x}b/2+ik_{y}\sqrt{3}b/2)\,]$,
$t_{2\mathbf{k}}=\gamma_{0}\exp[\,(ik_{x}b/2-ik_{y}\sqrt{3}b/2)\,]$,
and $t_{3\mathbf{k}}=\gamma_{0}\exp(\,-ik_{x}b\,)$. After specifying
the matrix elements, the zero-field Hamiltonian is expressed as the
following $2\times 2$ Hermitian matrix,
\begin{eqnarray}
    \left(\begin{array}{cc}
        0 & \sum_i t^{\ast}_{i\mathbf{k}}\\
        \sum_i t_{i\mathbf{k}} & 0
    \end{array}\right).
\end{eqnarray}
The Hamiltonian matrix can be solved analytically, and it generates
two energy dispersions represented by
$E_{\pm}(\mathbf{k})=\pm\gamma_0[1+4\cos^2{(\sqrt{3}k_y
b/2)}+4\cos{(\sqrt{3}k_y b/2)}\cos{(3bk_x/2)}]^{1/2}$. The
low-energy spectrum shows symmetry between two valleys (around K and
K'), where both have the identical light-cone structure described by
the relation $E_{\pm}(\mathbf{q})=\pm\upsilon_F|\mathbf{q}|$, with
light velocity $\upsilon_F\equiv3b\gamma_0/2$ and momentum
$\mathbf{q}$ measuring the difference
$\mathbf{q}=\mathbf{k}-\mathbf{k}_{K}$
($\mathbf{k}-\mathbf{k}_{K'}$) in wavevector $\mathbf{k}$ and corner
$\mathbf{k}_K$ ($\mathbf{k}_{K'}$) of the Brioullin zone.

Considering graphene exists in a 1D modulated potential $U(x)$ along
the armchair direction ($\hat{x}$), and the potential profile is
assumed to take the form $U(x)=V_0 \cos(2\pi x/l_E)$ with $V_0$
being the field strength. For convenience, the period, $l_E$, is
further designed to be a multiple of $3b$, with $l_E=3bR_E$. In
doing so, the periodicity caused by such a field is made
commensurate with the crystal potential of graphene itself, which
helps define a primitive cell. The rectangular primitive cell is
enlarged to include $4R_E$ carbon atoms denoted as $A_n$($B_n$) with
$n=1,2,\ldots,2R_E$ for A(B)-type carbon atoms (Fig. 1(a)). The
corresponding rectangular Brioullin zone shrinks to be $1/2R_E$ of
the original hexagonal Brioullin zone, and its dimension along the
modulated direction ($|k_x|\leq\pi/3bR_E$) is relatively shorter
than the other one ($|k_y|\leq\pi/\sqrt{3}b$), as shown in Fig.
1(b). The eigenfunction of a system is the superpostion of elements
in the basis composed of $4R_E$ Bloch functions,
$\{|\phi^{A_1}_{\mathbf{k}}\rangle,|\phi^{B_{1}}_{\mathbf{k}}\rangle,
|\phi^{A_2}_{\mathbf{k}}\rangle,|\phi^{B_{2}}_{\mathbf{k}}\rangle,\ldots
|\phi^{A_{2R_E-1}}_{\mathbf{k}}\rangle,|\phi^{B_{2R_E-1}}_{\mathbf{k}}\rangle,
|\phi^{A_{2R_E}}_{\mathbf{k}}\rangle,|\phi^{B_{2R_E}}_{\mathbf{k}}\rangle\}$,
and is represented as
\begin{eqnarray}
|\Psi_{\mathbf{k}}\rangle=\sum
C^{A_n}_{\mathbf{k}}|\Phi_{\mathbf{k}}^{A_n}\rangle+C^{B_n}_{\mathbf{k}}|\Phi^{B_n}_{\mathbf{k}}\rangle.
\end{eqnarray}
When the period is sufficiently large, the effects of the electric
potential on the off-diagonal matrix elements are negligible.
Meanwhile, the the diagonal matrix elements would become
\begin{eqnarray}
\langle\,\Phi^{A_n}_{\mathbf{k}}|H|\Phi^{A_n}_{\mathbf{k}}\rangle&=
V_0\cos{[(n-1)\pi/R_E]}\equiv U_n;\nonumber \\
\langle\,\Phi^{B_n}_{\mathbf{k}}|H|\Phi^{B_n}_{\mathbf{k}}\rangle&=
V_0\cos{[(n-2/3)\pi/R_E]}\equiv U_{n+1/3},
\end{eqnarray}
where $H=H_0+U$ is the total Hamiltonian. In effect, the assumption
made above is plausible. The period ($\geq 0.1$ $\mu$m) in a typical
experimental setup is still much longer than the extension of atomic
$2p_z$ orbital ($\approx 1$ $\AA$), so the atomic orbital feels
approximately a constant potential within its distribution. To
facilitate the calculation, the basis functions are further
rearranged in a specific sequence to obtain the band Hamiltonian
matrix,
$\{|\phi^{A_1}_{\mathbf{k}}\rangle,|\phi^{B_{2R_E}}_{\mathbf{k}}\rangle,
|\phi^{B_1}_{\mathbf{k}}\rangle,|\phi^{A_{2R_E}}_{\mathbf{k}}\rangle,\ldots
|\phi^{A_{R_E}}_{\mathbf{k}}\rangle,|\phi^{B_{R_E+1}}_{\mathbf{k}}\rangle,
|\phi^{B_{R_E}}_{\mathbf{k}}\rangle,|\phi^{A_{R_E+1}}_{\mathbf{k}}\rangle\}$.
The band Hamiltonian matrix of graphene in a modulated electric
potential along the armchair direction is written as
\begin{eqnarray}
\left(
  \begin{array}{cccccccc}
    U_1 & q & p^{\ast} & 0 & \ldots & \ldots & 0 & 0 \\
    q^{\ast} & U_{2R_E+1/3} & 0 & p & 0 & \ldots & \ldots & 0 \\
    p & 0 & U_{4/3} & 0 & q^{\ast} & 0 & \ldots & 0 \\
    0 & p^{\ast} & 0 & U_{2R_E} & 0 & q & 0 & 0 \\
    \vdots & \ddots & q & 0 & U_2 & \ddots & \ddots & 0 \\
    \vdots & \ldots & \ddots & q^{\ast} & \ddots & \ddots & 0 & p \\
    0 & \vdots& \vdots & \ddots & \ddots & 0 & U_{R_E+1/3} & q^{\ast} \\
    0 & 0 & 0 & 0 & 0 & p^{\ast} & q & U_{R_E+1} \\
  \end{array}
\right),
\end{eqnarray}
with $p\equiv t_{1\mathbf{k}}+t_{2\mathbf{k}}$ and $q\equiv
t_{3\mathbf{k}}$.

Graphene owns the anisotropic geometry so that the $\pi$-electronic
structure could depend on the modulation direction.  When the
modulation direction is changed into the zigzag direction
($\hat{x}$), the diagonal matrix element becomes
\begin{eqnarray}
\langle\,\Phi^{A_n}_{\mathbf{k}}|H|\Phi^{A_n}_{\mathbf{k}}\rangle=\langle\,\Phi^{B_n}_{\mathbf{k}}|H|\Phi^{B_n}_{\mathbf{k}}\rangle=
U_n.
\end{eqnarray}
In this case, the period is designed as $l'_E=\sqrt{3}bR_E$ for the
same reason mentioned earlier in the armchair case. The
corresponding rectangular Brioullin zone also shrinks to be $1/2R_E$
of the original hexagonal Brioullin zone, and its dimension is
characterized by $|k_x|\leq\pi/\sqrt{3}bR_E$ along the zigzag
direction and $|k_y|\leq\pi/3b$ along the other. The Hamiltonian
matrix can be easily constructed in a similar way to that in the
armchair case (not shown here).

In the following discussion, we will consider the situation that the
actual modulation period is about submicron length, which
corresponds to $R_E\approx 250$. This involves a process of
diagonalizing a very large Hamiltonian matrix, and it is solved
numerically to obtain the energy bands. Due to the resulting
unoccupied conduction bands ($E^{c}$) and occupied valence bands
($E^{v}$) being symmetric about the Fermi level ($E_F=0$), we only
discuss the former.

\section{Electronic Properties}

Before presenting the results under a modulated electric potential,
a brief review of the main features of low-energy bands of the
zero-field system is made. As mentioned earlier, the low-energy band
structure has a linear energy dispersion around K or K' of the
hexagon. However, it is not convenient to compare directly with the
coming case when a modulated field is present, because they are
presented in different Brillouin zones. To treat them on an equal
footing, the unit cell of the zero-field system is chosen to be
identical to the primitive cell of the system in a modulated
potential. That is, if the modulated field has a period $R_E$, the
unit cell is chosen to be $2R_E$ times its primitive cell. In this
way, all electronic states in the hexagonal Brillouin zone are
folded into a rectangular one. Notice that such a folding method
guarantees not to change any features of the original band
structure. Figure 1(c) shows the $k_y$-dependence of the low-energy
spectrum at $R_E=250$ for $k_x=0$ (black solid lines) and
$k_x=\pi/3bR_E$ (blue dotted lines). There exists a nondegenerate 1D
linear band and doubly degenerate 1D parabolic bands for $k_x=0$,
while they are purely doubly degenerate 1D parabolic bands for
$k_x=\pi/3bR_E$. The evolution of bands from $k_x=0$ to
$k_x=\pi/3bR_E$ fills the states between them, which reflects the 2D
characteristic of the light-cone structure. (This energy spectrum
corresponds to one of the two valleys containing the Dirac point K
now located at $k_x=0$ and $k_y=2\pi/3\sqrt{3}b$.) Above the Dirac
point, there exists a local minima for each 1D parabolic band.
Notice that these points should not be considered as band-edge
states. For these states, the $k_x$-dependent energy dispersion is
linear with nonzero first derivative, so they cannot be treated as
critical points. The band structure is in all respects the same as
one before folding the zone. Although this alternative
representation of band structure somewhat complicates the
explanation, it is advantageous for the following discussion to
compare the zero-field system and the system in a modulated electric
potential.

A modulated electric potential has a strong effect on the energy
dispersions, state degeneracy, and band-edge states. The
$k_y$-dependent conduction bands for $k_x=0$ (black solid lines) and
$k_x=\pi/3bR_E$ (blue dotted lines) are shown in Fig. 2(a) for the
modulated electric potential along the armchair direction with the
period $R_E=250$ ($\approx100$ nm) and the strength $V_0=0.025$
$\gamma_0$. Moreover, the potential mainly affects the structure of
some low 1D bands. The original doubly degenerate parabolic bands
are split. The energy dispersions around the
$k_y=2\pi/3\sqrt{3}b\equiv k_y^{K}$ are strongly deformed, and
induce several band-edge states which are saddle points in the
energy-momentum space. These band-edge states always appear in pairs
at two sides of $k_y^{K}$, and two band-edge states is each pair
might have small energy difference. Far away from the $k_y^{K}$, the
spectrum is almost linear for the $k_y$ dependence, while it becomes
dispersionless for the $k_x$ dependence, which can been seen in Fig.
2(a) where the bands for $k_x=0$ and $k_x=\pi/3bR_E$ have identical
dispersion. It is important to note that there exist more Fermi
momentum states $k_{F}$'s. The energy dispersions near $k_{F}$'s are
linear for the $k_{y}$-dependence, while they are completely flat
for the $k_{x}$-dependence. The dispersionless feature means that
the number of the Fermi-momentum states is infinite, which sharply
contrasts with the original two Fermi-momentum states or Dirac
points in zero-field graphene. Besides, the non-zero measure of the
Fermi surface indicates finite value of the DOS at the Fermi level.

The aforementioned features might rely on the strength, period, and
direction of a modulated electric potential. The modulation effects
are even enhanced when field is strengthened, illustrated in the
Fig. 2(b). When the strength increases, more low 1D bands are
affected. The energy bands are further deformed, and more band-edge
states are created. In addition, the number of Fermi-momentum states
increases almost linearly with the field strength. For the case in
Fig. 2(b), the number of Fermi-momentum states is approximately
double that in Fig. 2(a). The modulation period can also affect the
numbers of band-edge states and Fermi-momentum states, as shown in
Fig. 2(c). When the period increases, more band-edge states are
created. The number of Fermi-momentum states remains almost
unchanged in the long-period regime, whereas there is a drastic
change in the small-period regime. The former can be understood by
taking into consideration both the accompanying reduced $k_x$ length
and increased $k_F$'s associated with different $k_y$'s in
accordance with the increasing period, while the latter will be
discussed later by calculating the density of states at the Fermi
level. The low-energy bands are, on the other hand, not sensitive to
the direction of a modulated electric potential. Figure 2(d), for
example, presents the bands for the potential modulated along the
zigzag direction.

The essential features of electronic structure are directly
reflected on the density of states, and defined as
\begin{eqnarray}
D(\omega \,)=\sum_{\sigma
\,,h=c,v}\int_{1stBZ}{\frac{dk_{x}dk_{y}}{(2\pi
\,)^{2}}}{\frac{\Gamma }{\pi }}{\frac{1}{[\omega
-E^{h}(k_{x},k_{y})]^{2}+\Gamma ^{2}}}\,.
\end{eqnarray}
Because of the linear energy dispersion, the low-frequency DOS
without fields is proportional to $\omega $, as shown by dotted line
in Fig. 3(a), and it has no special structures. The vanishing DOS at
$E_{F}=0$ indicates that graphene is a zero-gap semiconductor. A
modulated electric potential can lead to many prominent peaks and a
finite DOS at $\omega =0$ (Figs. 3(a)-3(d)). The peak structures
come from the band-edge states of parabolic bands (Figs. 2(a)-2(d)).
The frequency, number, and height of peaks are sensitive to the
changes in the field strength and period. The DOS at the Fermi level
means that there are free carriers, so graphene becomes a semimetal
in the presence of a modulated electric field. The value of DOS at
$\omega=0$ grows as the field strength increases (Figs. 3(a) and
3(b)). However, it does not change significantly as the period is
varied (Fig. 3(c)). Moreover, the low-frequency DOS is unaltered
when the modulation direction changes (Fig. 3(d)), which reflects
the isotropic symmetry of graphene in low-energy spectrum.

The singularities in the DOS might cause special structures in some
physical quantities, for example, giving rise to the strong
absorption peaks in optical measurement, so it is worthy to
investigate their properties in detail. The peak height, number and
frequency are dominated by the strength and period of the potential
(Figs. 3(a)-3(c)). The peak height is enhanced by the increasing
strength, whereas it is reduced by the increasing period. The peak
number is increased by both the increasing strength and period. The
relations between of the peak frequency ($\omega_{be}$) and the
field condition are elaborated through examining the first four
prominent peaks, as shown in the Fig. 4(a) for the strength
dependence and Fig. 4(b) for the period dependence. From these, it
is observed that the peak frequency weakly depends on the strength,
while declines and presents somewhat oscillatory behavior as the
period increases.

The relationship between the DOS at $\omega=0$ and the field
strength and that between the field strength and the period deserve
a closer investigation. Referring to Fig. 5(a), $D(\omega=0)$ shows
a nearly linear variation with the field strength. This relation
just reflects the fact that the number of Fermi-momentum states is
approximately proportional to field strength. On the other hand, the
period dependence of $D(\omega=0)$ exhibits different features in
short- and long-period regimes, as shown in Fig. 5(b). The two
regions are distinguished by a threshold period $R_{th}$, which
happens at $R_E\approx 100$ for $V_0=0.025$ $\gamma_0$ ( black
triangles in Fig. 5(b)) and becomes shorter as the field strength
gets stronger ( red circles in Fig. 5(b)). For $R_E\ll R_{th}$,
$D(\omega=0)$ remains zero as $R_E$ increases from zero, which
implies that the 2D light-cone structure is not affected at such a
short period. As $R_E$ is increased to cross $R_{th}$, $D(\omega=0)$
quickly grows and saturates to a definite value, with some
fluctuations in the long-period regime. If one consider a realistic
experimental system, the typical modulation period will be about
submicron or even longer. The $D(\omega=0)$ then would fall into the
long-period regime, so its value does not change significantly as
the period is varied. Therefore, this feature should be very robust
even there exist some inevitable period fluctuations coming from
processing devices or experimental setups.

These interesting features, including field induced band-edge states
and the semimetallic behavior of graphene, caused by the modulated
electric potential are the main results in this work. Nevertheless,
the semimetallic behavior, the most important feature, was not
obtained in a recent study \cite{mod-4}. The study considered
graphene subjected to a square-wave like modulated potential,
concluding that the 2D light-cone structure is still preserved.
Although the conclusion seems to contradict with our results given
here, a in-depth examination shows that this is not the case. The
main discrepancy lies in the fact that the modulation period used
(20 nm $\approx R_E=50$) in their calculation is too short to
achieve the threshold ($R_{th}$), so that the semimetallic behavior
was not observed. To clarify this point, the dependence of
$D(\omega=0)$ on the field strength and the modulation period for a
square-wave potential is calculated here, as shown by green squares
in Figs. 5(a) and 5(b) respectively. $D(\omega=0)$ in Fig. 5(a)
exhibits a linear relation with the field strength, which is similar
to the cosine-wave potential. The threshold period in this case is
$R_{th}\approx 100$ (Fig. 5(b)), which is close to that in the
cosine-wave potential. As a result, from Fig. 5(b), it
self-evidently explains the reason why the semimetallic behavior is
not observed by other authors solely because that $R_E$ taken in
their calculation is less than $R_{th}$.

\section{Concluding Remarks}

In summary, the tight-binding model is used to investigate the
effects of a modulated electric potential on the $\pi $-electronic
structure of graphene. The low-energy electronic properties are
mainly dominated by field strength and period. However, they are not
sensitive to the field direction, which is due to the isotropy of
the original low-energy bands. A modulated electric potential
drastically changes the energy dispersions, state degeneracy, and
band-edge states.  The density of states exhibits a lot of prominent
peaks. These structures are mainly determined by the energy
dispersions and the band-edge states. Notice that there are infinite
Fermi-momentum states. The finite DOS at the Fermi level indicates
the existence of free carriers. The semiconducting graphene becomes
semimetallic by applying a modulated electric potential, as long as
the period surpasses a threshold period ($R_E\gg R_{th}$). Different
nature of the low-energy excitations for graphene in the presence of
a modulated potential is expected to give rise intriguing physical
properties. For instance, the free electrons are expected to cause
the low-frequency plasmon. The theoretical predictions could be
tested by the experimental measurements on the energy loss spectra.

\begin{acknowledgments}
This work was supported by NSC and NCTS of Taiwan, under the Grant
Nos. NSC 96-2112-M-006-002.
\end{acknowledgments}

\bibliography{modE}

\newpage

\begin{figure}[p]
\begin{center}\leavevmode
\includegraphics[width=0.9\linewidth]{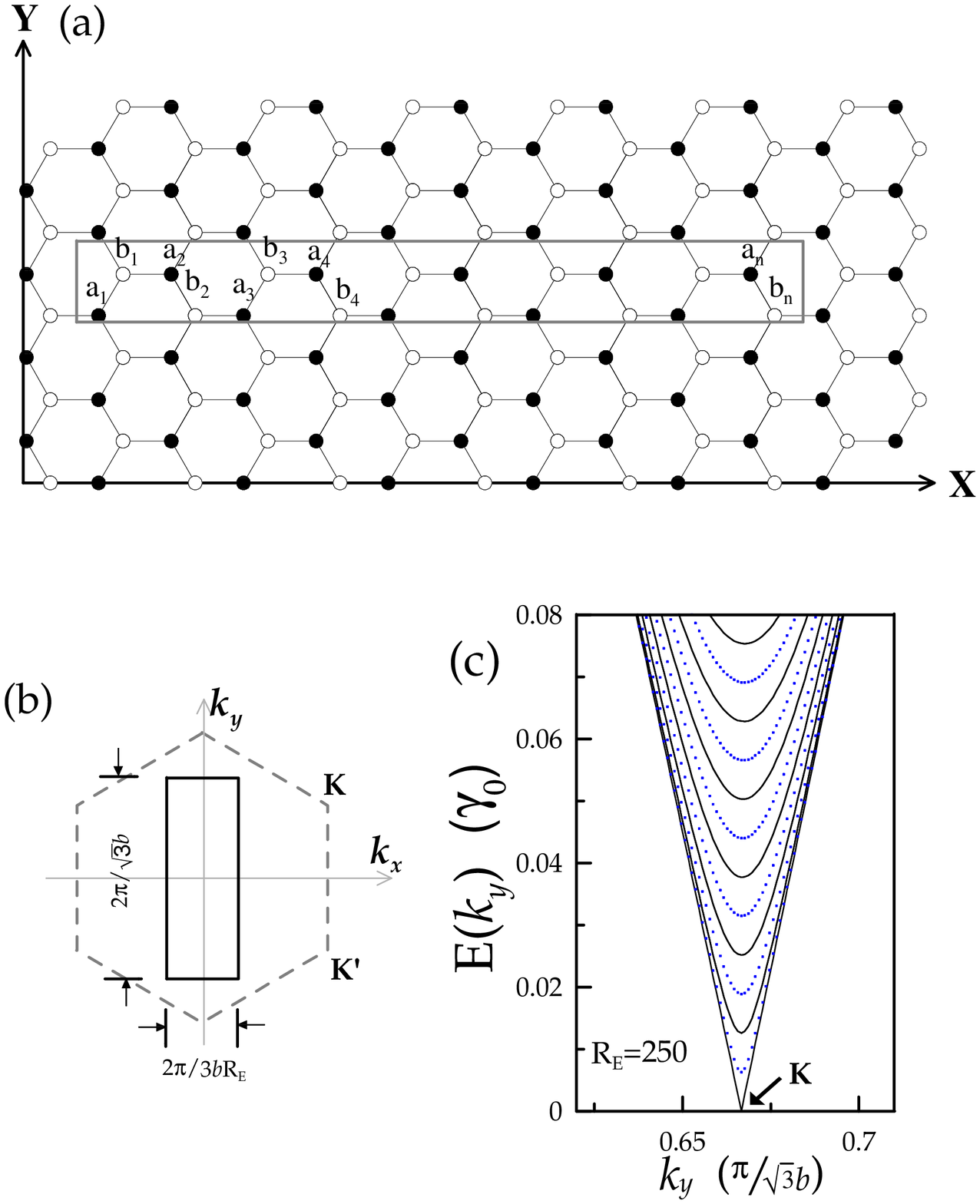}
\caption{(a) The rectangular primitive cell of graphene in a
modulated electric potential along the armchair direction
($\hat{x}$). The rectangular first Brillouin zone is shown in (b),
and the hexagon is that without external fields. The $k_y$-dependent
low-energy bands without any fields at $R_E=250$ is shown in (c) for
$k_x=0$ (black solid lines) and $k_x=2\pi/3bR_E$ (blue dotted
lines).}
\end{center}
\end{figure}

\begin{figure}[p]
\begin{center}\leavevmode
\includegraphics[width=0.8\linewidth]{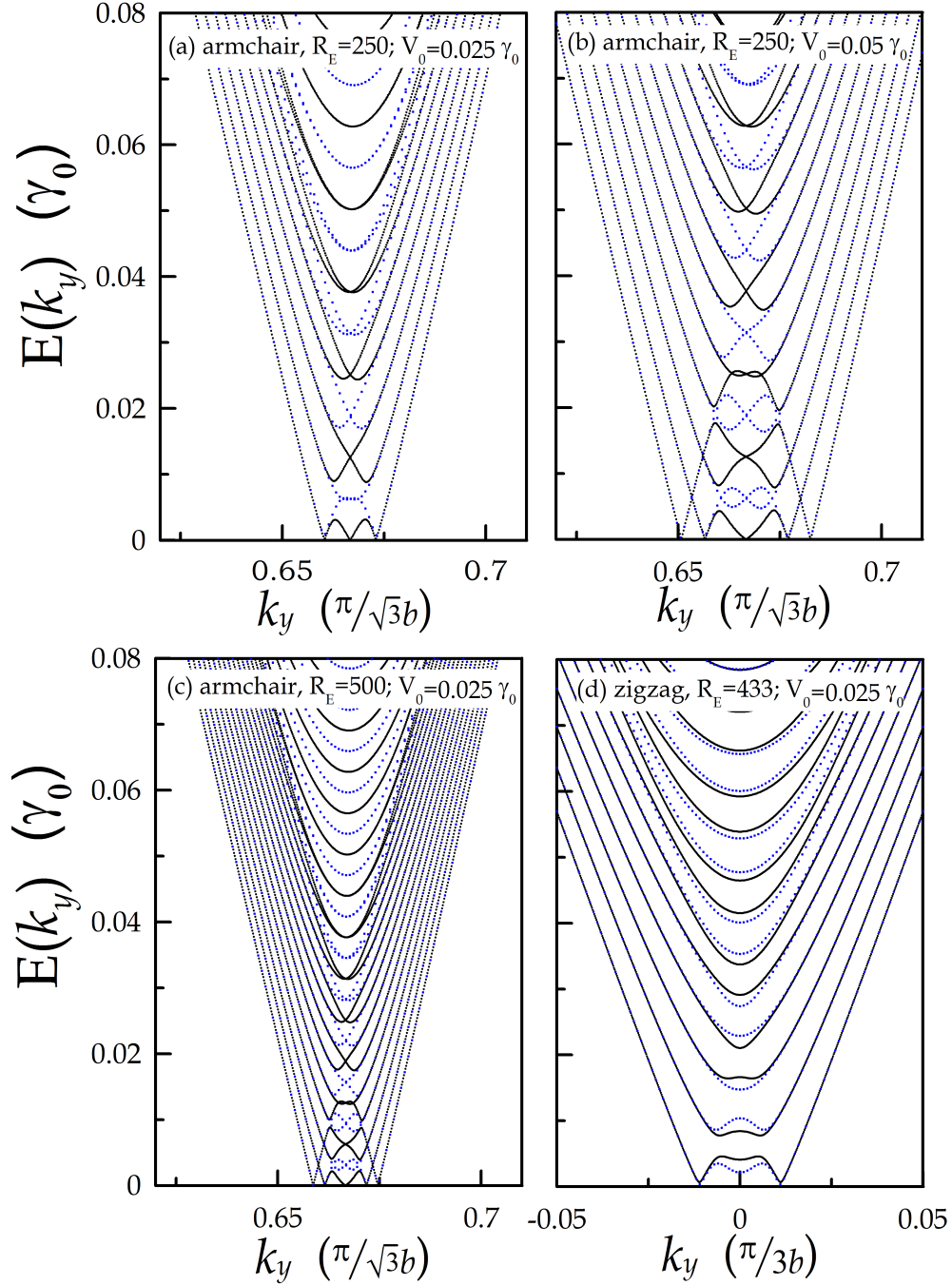}
\caption{The $k_y$-dependent low-energy bands in the modulated field
along armchair direction with $V_0=0.025$ $\gamma_0$ and $R_E=250$
for $k_x=0$ (black solid lines) and $k_x=2\pi/3bR_E$ (blue dotted
lines), those in a stronger field with $V_0=0.05$ $\gamma_0$, and
those in the field of a longer period $R_E=500$ are respectively
shown in (a), (b), and (c). (d) is the $k_y$-dependence of
low-energy bands at $V_0=0.025$ $\gamma_0$, $R_E=433$ with the
potential modulating along the zigzag direction for $k_x=0$ (black
solid lines) and $k_x=2\pi/\sqrt{3}bR_E$ (blue dotted lines). Notice
that the actual periods in (a) and (d) are almost of equal length,
and that the units of $k_y$ are different for armchair
($\pi/\sqrt{3}b$) and zigzag ($\pi/3b$) direction.}
\end{center}
\end{figure}

\begin{figure}[p]
\begin{center}\leavevmode
\includegraphics[width=0.9\linewidth]{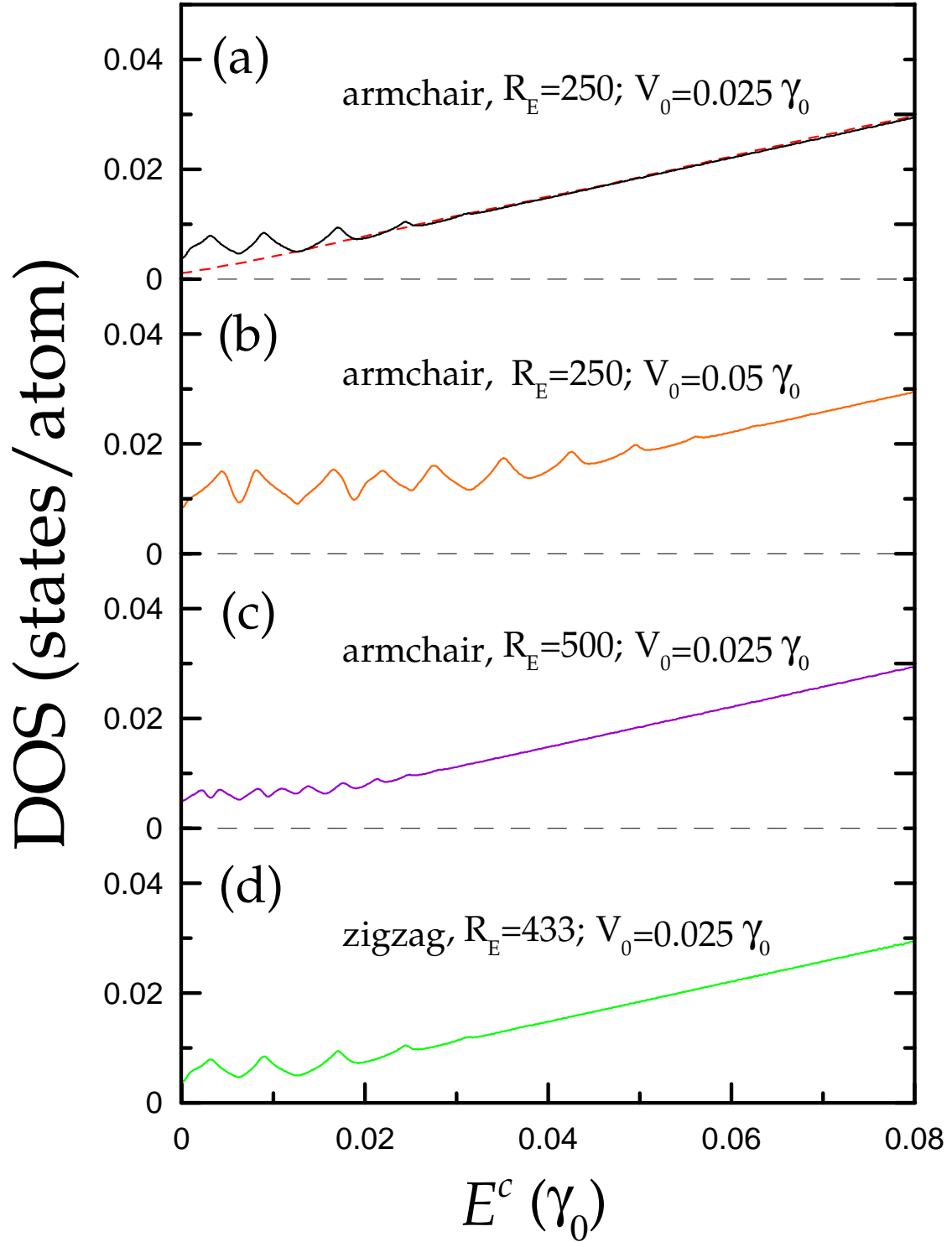}
\caption{The low-frequency density of states for the potential
modulated along armchair direction at (a) $V_0=0.025$ $\gamma_0$,
$R_E=250$; (b) $V_0=0.05$ $\gamma_0$, $R_E=250$; (c) $V_0=0.025$
$\gamma_0$, $R_E=500$; (d) along the zigzag direction at
($V_0=0.025$ $\gamma_0$, $R_E=433$). The dashed line in (a) is for
the zero-field spectrum.}
\end{center}
\end{figure}

\begin{figure}[p]
\begin{center}\leavevmode
\includegraphics[width=0.9\linewidth]{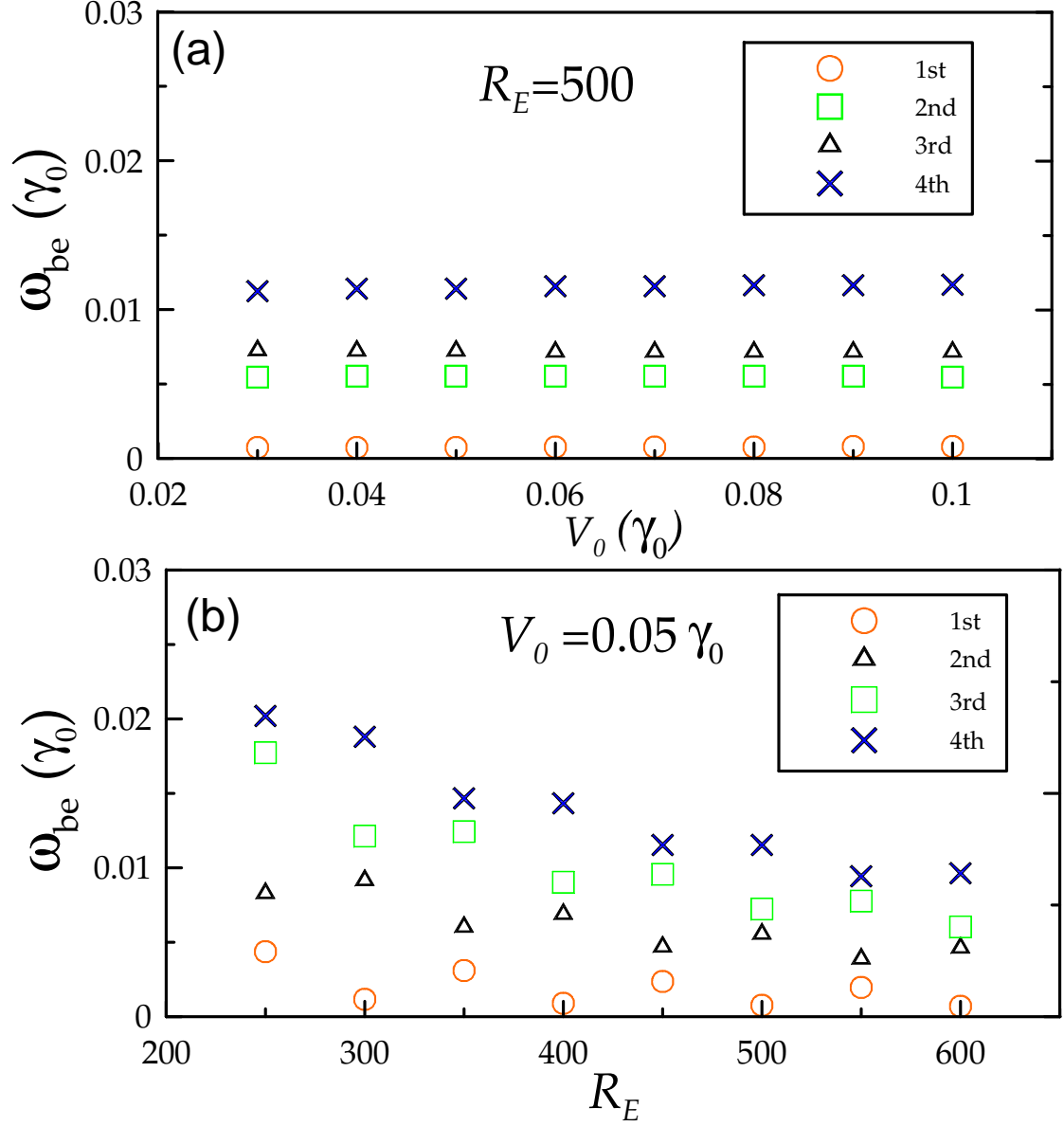}
\caption{Energies of the first four prominent peaks versus (a) the
field strength at $R_E=500$ and (b) the modulation period at
$V_0=0.05$ $\gamma_0$.}
\end{center}
\end{figure}

\begin{figure}[p]
\begin{center}\leavevmode
\includegraphics[width=0.9\linewidth]{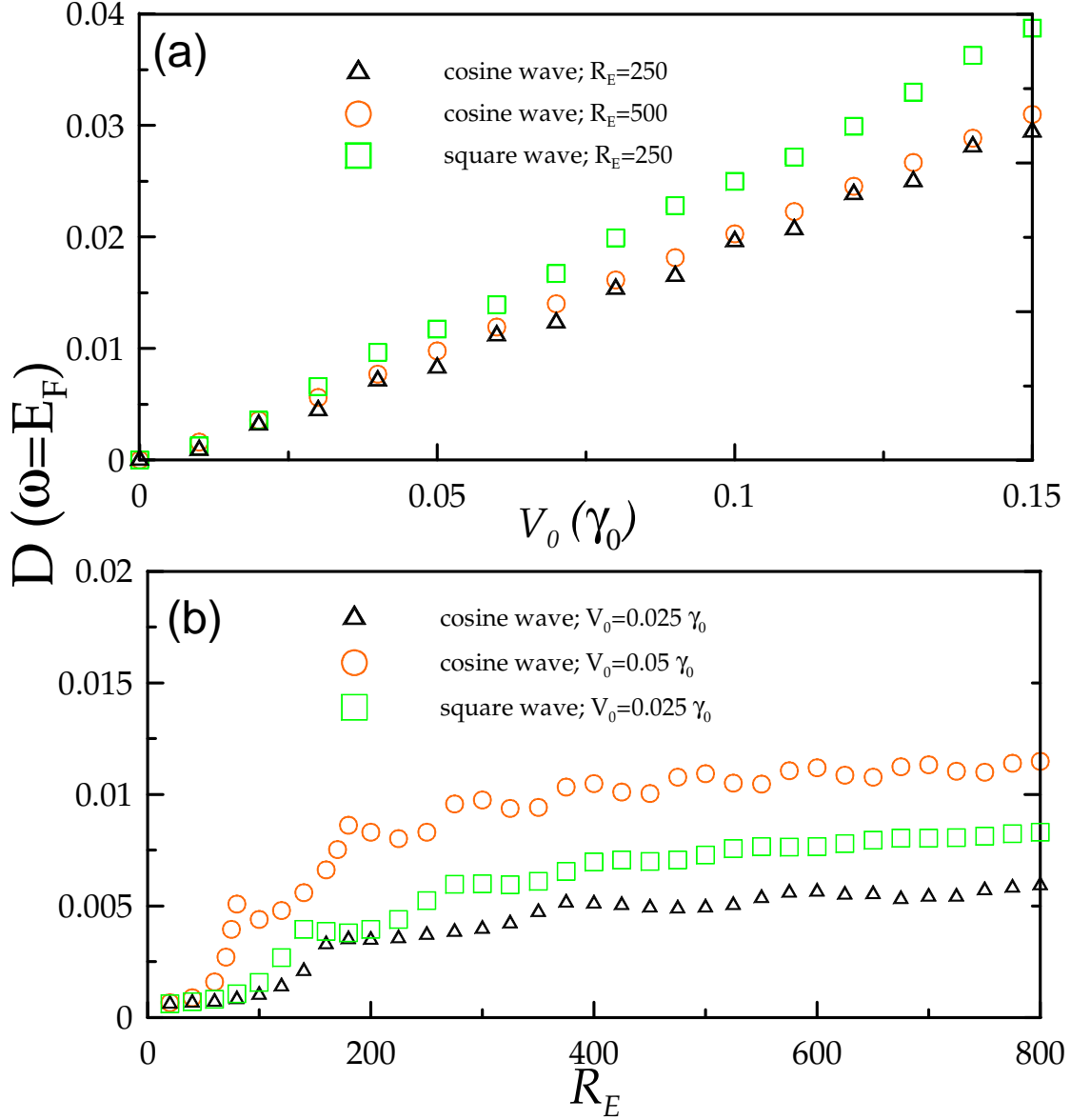}
\caption{The density of states at the Fermi level versus (a) the
field strength at $R_E=250$ and $500$ ($R_E=250$) for the modulated
potential of cosine-wave (square-wave) profile, and versus (b) the
modulation period at $V_0=0.025$ and $0.05$ $\gamma_0$ ($V_0=0.025$
$\gamma_0$) for that of the cosine-wave (square-wave) profile.}
\end{center}
\end{figure}

\newpage
\end{document}